\documentclass[12pt]{article}
\usepackage{times}
\usepackage{geometry}
\usepackage[utf8]{inputenc}
\usepackage{enumitem,amssymb}
\usepackage{ragged2e}
\usepackage{graphicx}
\newlist{thematic}{itemize}{8}
\setlist[thematic]{label=$\square$}
\usepackage{pifont}

\newcommand{\hi}{\textsc{HI}}

\newcommand{\mhi}{$M_{HI}$}
\newcommand{\nhi}{$N_{HI}$}

\newcommand{\cmsq}{cm$^{-2}$}

\begin{document}
\raggedright
\huge
Astro2020 Science White Paper \linebreak

Completing the Hydrogen Census in the Circumgalactic Medium at z$\sim$0 \linebreak
\normalsize

\noindent \textbf{Thematic Areas:} \hspace*{60pt} $\square$ Planetary Systems \hspace*{10pt} $\square$ Star and Planet Formation \hspace*{20pt}\linebreak
$\square$ Formation and Evolution of Compact Objects \hspace*{31pt} $\square$ Cosmology and Fundamental Physics \linebreak
  $\square$  Stars and Stellar Evolution \hspace*{1pt} $\square$ Resolved Stellar Populations and their Environments \hspace*{40pt} \linebreak
  $\boxtimes$    Galaxy Evolution   \hspace*{45pt} $\square$             Multi-Messenger Astronomy and Astrophysics \hspace*{65pt} \linebreak
  
\textbf{Principal Author:}

Name:  D.J. Pisano
 \linebreak						
Institution:  Dept. of Physics \& Astronomy and the Gravitational Wave and Cosmology Center, West Virginia University
 \linebreak
Email: djpisano@mail.wvu.edu
 \linebreak
Phone:  +1-304-293-4886
 \linebreak
 
\textbf{Co-authors:} A. Fox, D. French (STScI), J.C. Howk, N. Lehner (University of Notre Dame), F.J. Lockman (Green Bank Observatory), K. Jones (NAIC)
\linebreak

\textbf{Abstract:}

Over the past decade, Lyman-$\alpha$ and metal line absorption observations have established the ubiquity of a gas-rich
circumgalactic medium (CGM) around star-forming galaxies at z$\sim$0.2 potentially tracing half of the missing baryonic
mass within galaxy halos.  Unfortunately, these observations only provide a statistical measure of the gas in the CGM and 
do not constrain the spatial distribution and kinematics of the gas.  Furthermore, we have limited sensitivity to Lyman-$\alpha$ 
at z$\sim$0 with existing instruments.  As such, we remain ignorant of how this gas may flow from the CGM onto the disks 
of galaxies where it can fuel ongoing star-formation in the present day.  Fortunately, 21-cm \hi\ observations with radio 
telescopes can map \hi\ emission providing both spatial and kinematic information for the CGM in galaxies at z$=$0.  Observations
with phased array feeds, radio cameras, on single-dish telescopes yield unmatched surface brightness sensitivity and
survey speed.  These observations can complete the census of \hi\ in the CGM below N$_{HI}\lesssim$10$^{17}$cm$^{-2}$ and
constrain how gas accretion is proceeding in the local universe, particularly when used in concert with UV absorption line data.

\vspace{12pt}

\pagebreak

\section{Background}

Great strides have been made in understanding the nature and evolution of galaxies over the past fifty years.  We know that 
galaxies assemble their mass in a hierarchical manner by accreting smaller galaxies with their associated stars and dark matter and 
merging with other galaxies in a process that continues to the present day \cite{press74}.  It is still unknown, however, 
how dark matter halos, and the galaxies contained therein, accrete the gas that they need to continue to form stars to the present 
day.  Current theories suggest that there are three ways a galaxy can accrete gas.  The most straight-forward is through the 
accretion of a satellite galaxy as part of the hierarchical assembly of a galaxy, but such gas-rich satellites are neither abundant enough nor contain enough
gas to sustain star formation\cite{diTeodoro14}.  Alternatively, gas can flow onto galaxies in either a hot (T$\sim$10$^6$K) or cold (T$\lesssim
$10$^5$K) phase \cite{keres05,keres09}.  The hot mode involves gas falling onto galaxies in a quasi-spherical mode 
and is expected to be dominant for high mass galaxies in higher density environments in the present day.  In contrast, the cold 
mode is more filamentary in nature and should dominate at high redshift and for low mass galaxies in lower density environments.  
While simulations disagree on the amount and exact phase of this accretion \cite{joung12,nelson13}, they all
agree that accretion from the intergalactic medium through the circumgalactic medium (CGM) and onto galaxy disks should still
be occurring today.

There is certainly evidence for ongoing accretion onto galaxies in the form of discrete, cold \hi\ clouds \cite{sancisi08,fox17}, which is likely 
tracing a larger, warm-hot ionized reservoir of gas \cite{lehner11}.  Further evidence for cold accretion comes from Lyman-limit absorption 
systems with low metallicities associated with nearby galaxies \cite{stocke10,ribaudo11}.  Absorption line studies, however, can only provide 
line-of-sight information and do not yield a complete picture of the gas through the CGM of individual galaxies.  {\it In order to understand 
how gas is accreted onto galaxies in the present day, comprehensive surveys in both emission and absorption of the CGM is required.}

\section{Current Observations of the CGM}

To date, most of the exploration of the CGM has come through UV absorption line studies.  
The COS-HALOS project \cite{tumlinson13,werk14} has used background quasars to study the 
Lyman-$\alpha$ absorption in the halos of low redshift galaxies.  The project has found that \hi\ 
absorption at \nhi$\gtrsim$10$^{14}$\cmsq\ is ubiquitous out to 150 kpc for star-forming galaxies and present in 75\% of
passive galaxies as well \cite{tumlinson13}.  This cool CGM gas represents 25\%-45\% of the total baryon mass
within the virial radius of the galaxy \cite{werk14, tumlinson17}.  Unfortunately, above \nhi$\sim$10$^{16}$\cmsq\ saturation of
absorption lines makes it difficult to get an accurate measure of \nhi; these are the Lyman Limit Systems.  While
below \nhi$\sim$10$^{16-17}$\cmsq, \hi\ absorption is common, particularly in the intergalactic medium, it has been impossible
to image in 21-cm \hi\ emission to date.  Obtaining deeper \hi\ emission observations is the only way forward: while Lyman-$\alpha$ 
absorption observations can reach very low column densities, their pencil-beam nature make it extremely difficult to reconstruct the 
full gas distribution and its kinematics.  Furthermore, such observations are needed to measure \nhi\ at these column densities so 
that metallicities can be accurately determined.  Such metallicity measurements are key to understanding if the gas in the CGM is  
infalling, pristine gas or enriched outflows.

Over the past decade, 21-cm \hi\ emission observations have yielded great insights into the nature of the CGM around nearby 
galaxies.   Single-dish and interferometric observations have revealed both discrete \hi\ clouds as well as diffuse \hi\ structures that
are related to previously unknown dwarf galaxies, tidal interactions or accretion events 
\cite{hess09,pisano11,mihos12,pisano14,deblok14,pingel18,sorgho19}.  While such observations detect less than 10\% more \mhi, this emission is tracing the more massive, dominant, ionized gas reservoir in the CGM of these galaxies \cite{lehner11}.  A prime example of this are the 
extensive \hi\ surveys of the M~31's CGM.  \cite{braun04} discovered a \hi\ bridge between M~31 and M~33 with \nhi$\gtrsim$10$^{17}$cm
$^{-2}$ that they attributed to the cosmic web (Figure~\ref{fig:m31}).  Higher resolution observations with the Green Bank Telescope (GBT) 
have shown that this diffuse structure is actually comprised of discrete, higher \nhi\ clouds \cite{wolfe13,wolfe16}.  At these \hi\ column densities the gas clouds are mostly ionized \cite{maloney93}, but they allow us to trace the morphology and kinematics of this reservoir.  Furthermore, 
the CGM of M~31 itself is quite clumpy with \hi\ covering fractions below 5\% at \nhi$\sim$4$\times$10$^{17}$\cmsq\ \cite{howk17}.  As seen in Figure~2, these results are consistent with simulations, but are significantly lower than what was found from COS-HALOS \cite{prochaska17}.  This could be due to the unique
properties of M~31 or represent the evolution of the CGM since z$\sim$0.2, where most COS-HALOS galaxies reside.  These results demonstrate the need for high spatial angular resolution \hi\ surveys with excellent surface brightness sensitivity that can only be provided by large single-dish telescopes in concert with interferometers.

\section{The role of 21-cm \hi\ observations in the next decade}

\begin{figure}[!hb]
\begin{center}
\includegraphics[width=0.9\textwidth]{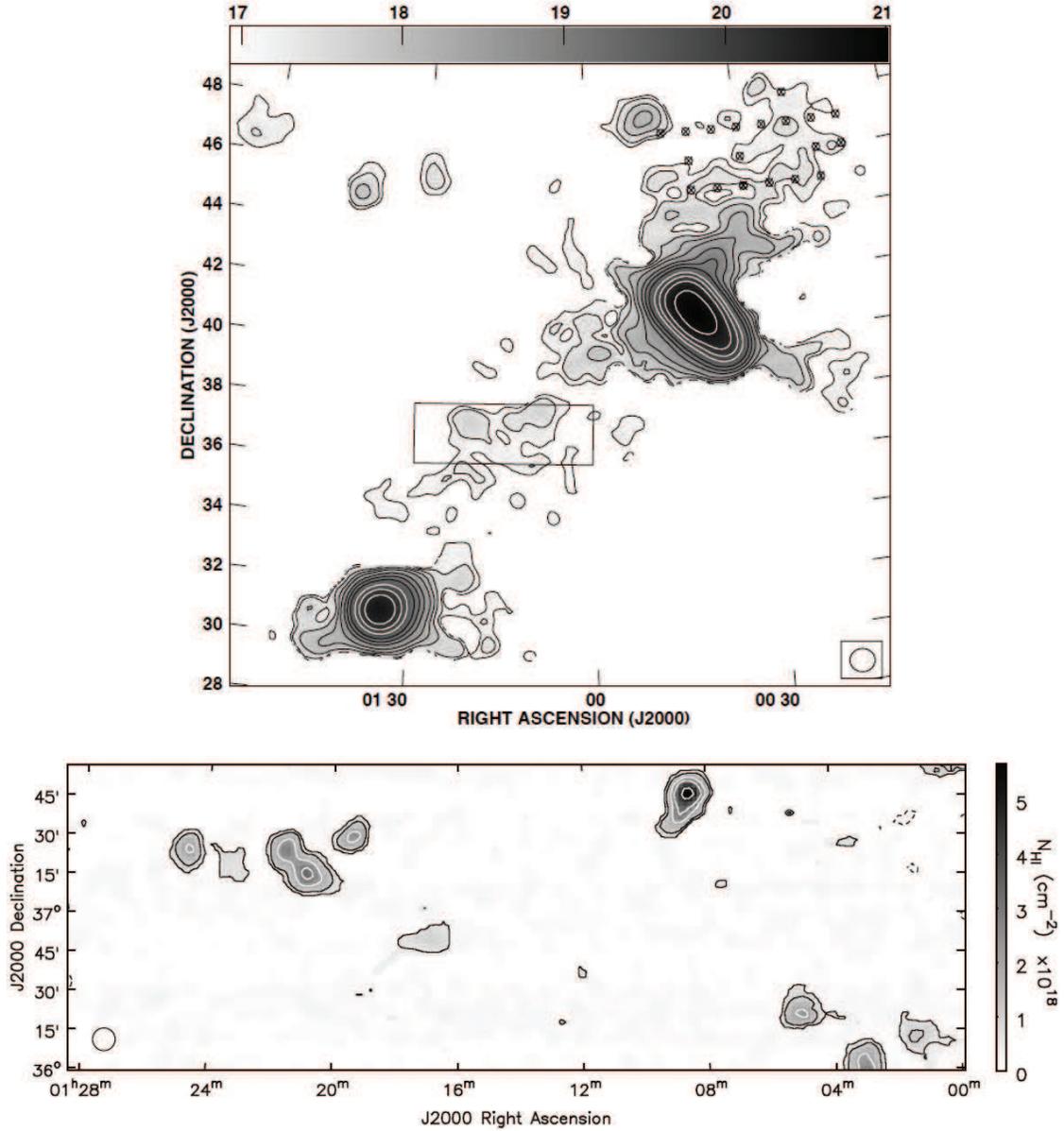}
\end{center}
\caption{Top:  A map of the total \hi\ emission associated with M~33 (lower left) and M~31 (upper right) from \cite{braun04}.  
The contours are at log \nhi$=$17.0, 17.3, 17.7, 18.0, 18.3, 18.7, 19.0, 19.3, 19.7, 20.0, 20.3, and 20.7 [\cmsq].  The beamsize
is shown in the lower right of the panel.  The box shows the region mapped by \cite{wolfe16} with the GBT.  Bottom:  
The GBT \hi\ map from \cite{wolfe16}.  The contours are at -1, 1, 2, 4, 6, and 10 times 5$\times$10$^{17}$\cmsq.  The beam 
size is the circle in the lower left of the image.  Note that the \hi\ structures detected by \cite{braun04} are revealed to be much smaller, 
higher-\nhi\ features by \cite{wolfe16}, illustrating the critical importance of resolution and sensitivity. \label{fig:m31}}
\end{figure}

\begin{figure}
\begin{center}
\includegraphics[width=0.9\textwidth]{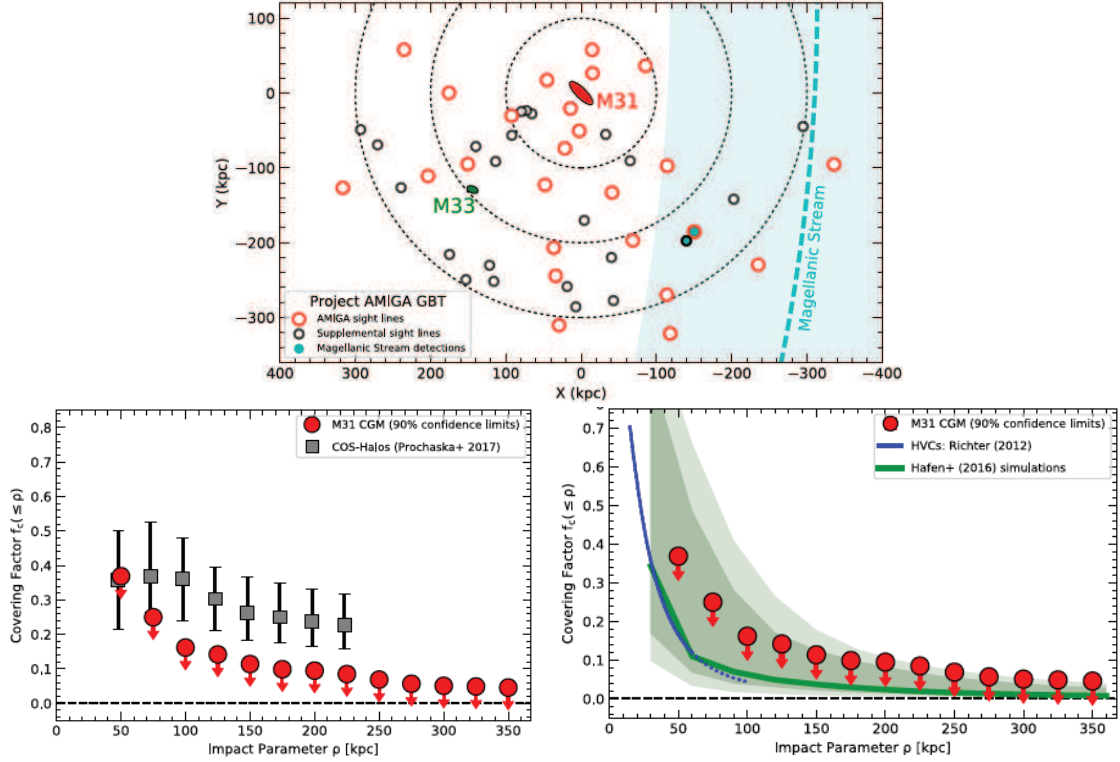}
\end{center}
\caption{Top:  Locations of the AMIGA \cite{howk17} GBT pointings relative to M31 and M33, where the axes are labeled with the impact parameter from the center of M31.  Aside from two filled circles, representing detections of \hi\ associated with the Magellanic Stream, all AMIGA observations yielded non-detections.  Bottom:  The cumulative covering fraction as a function of impact parameter of \hi\ emission with log(\nhi)$\ge$17.6 as compared to absorption line data from COS-HALOS \cite{prochaska17} (left), and high-velocity
clouds \cite{richter12} and simulations from \cite{hafen17}.  While the AMIGA data \cite{howk17} are consistent with simulations and
the Milky Way high-velocity clouds, they are inconsistent with the COS-HALOS data.  \hi\ and UV absorption line observations of more nearby galaxies will
help shed light on this discrepancy.  \label{fig:amiga}}
\end{figure}

As can be seen from Figures~\ref{fig:m31} and ~\ref{fig:amiga}, 21-cm \hi\ observations of the CGM of even a single galaxy provide
direct measurements of \nhi, independent of the optical depth of the gas, as well as the detailed morphology and kinematics of that
gas.  When combined with UV absorption line data, we can determine the metallicity of the CGM, which provides a strong constraint
on the origin of the gas.  To date, however, such deep (\nhi$\sim$10$^{17}$\cmsq) \hi\ emission observations have been limited to 
M~31.  The filaments of gas associated with cold accretion are expected to have widths up to $\lesssim$25 kpc \cite{popping09}, so 
spatial resolution is needed.  M~31 is close enough that single-dish telescopes, like the GBT, can spatially resolve \hi\ structures down to $\sim$2 kpc in its 
CGM mitigating the effects of beam dilution.  The GBT should be capable of resolving such filaments out to D$\sim$10 Mpc, while Arecibo could do so out to 
$\sim$30 Mpc.  Interferometers provide better resolution and have excellent \mhi\ sensitivity, but lack the surface brightness sensitivity needed to detect such low-\nhi\ emission.  These observations are valuable for
detecting \hi\ clouds around galaxies as demonstrated by HALOGAS \cite{heald11}.    Still to recover \hi\ emission at low \nhi, as
well as from compact sources, we need both single-dish and interferometric observations.  If we wish to study how the properties
of the CGM vary with galaxy mass and environment, we need to extend these observations to larger samples of galaxies beyond M~31.

In the next decade it will be possible to use existing single-dish telescopes, such as the GBT and Arecibo, outfitted with new
phased array feeds (PAFs), or radio cameras, FLAG \cite{roshi18} and ALPACA, to make sensitive (\nhi$\lesssim$10$^{17}$\cmsq) surveys
covering the entire dark matter halo of $\sim$100 galaxies within $\sim$20 Mpc spanning a range of masses and environments.  Due to the 
dramatic improvements in survey speed from PAFs, astronomers will be able to probe more diffuse gas around more galaxies.  These data will 
be capable of resolving filamentary structures associated with cold accretion and will provide
a complete census of the \hi\ content of the CGM of these galaxies.  By measuring the metallicity of these features and modeling 
their kinematics, we will be able to identify ongoing accretion events.  When combined with interferometer data, from the ngVLA 
for example, we will be able to trace accretion from the CGM directly on to the disks of galaxies.

The insight provided by such a survey will not be achievable without single-dish radio telescopes, as even the Square Kilometer 
Array will not achieve such excellent \nhi\ sensitivity.  While FAST in China will have better sensitivity and resolution than Arecibo or
 the GBT, the large field of view achieved with PAFs on these telescopes will result in faster survey speeds.

 Future UV studies with large-aperture space telescopes also have an important role to play in CGM studies in the next decade and beyond. UV facilities with multiplexing ability and higher sensitivity than Hubble/COS could be used to measure Lyman-$\alpha$\, in multiple QSO sightlines in a given galaxy halo. This would allow studies of spatial variation, kinematic structure, and covering fraction of \hi\ within individual halos. These observations would reach very low \hi column densities (10$^{13}$\cmsq) and hence complement the 21-cm radio observations that probe higher \hi\ column densities

\clearpage
\pagebreak

\bibliography{references.bib}{}  % For BibTex\end{document}

\end{document}